# The electronic structure of epitaxially stabilized 5$d$ perovskite Ca$_{1-x}$Sr$_x$IrO$_3$ ($x$ = 0, 0.5, and 1) thin films: the role of strong spin-orbit coupling

S. Y. Jang,[1] H. S. Kim,[2] S. J. Moon,[1] W. S. Choi,[1] B. C. Jeon,[1] J. Yu,[2] and T. W. Noh[1,*]

[1]ReCOE, Department of Physics and Astronomy, Seoul National University, Seoul 151-747, Korea

[2]CSCMR, Department of Physics and Astronomy, Seoul National University, Seoul 151-747, Korea

We have investigated the electronic structure of meta-stable perovskite Ca$_{1-x}$Sr$_x$IrO$_3$ ($x$ = 0, 0.5, and 1) thin films using transport measurements, optical spectroscopy, and first-principles calculations. We artificially fabricated the perovskite phase of Ca$_{1-x}$Sr$_x$IrO$_3$, which has a hexagonal or post perovskite crystal structure in bulk form, by growing epitaxial thin films on perovskite GdScO$_3$ substrates using epi-stabilization technique. The transport properties of the perovskite Ca$_{1-x}$Sr$_x$IrO$_3$ films systematically changed from nearly insulating (or semi-metallic) for $x$ = 0 to bad metallic for $x$ = 1. Due to the extended wavefunctions, 5$d$ electrons are usually delocalized. However, the strong spin-orbit coupling in Ca$_{1-x}$Sr$_x$IrO$_3$ results in the formation of effective total angular momentum $J_{eff}$ = 1/2 and 3/2 states, which puts Ca$_{1-x}$Sr$_x$IrO$_3$ in the vicinity of a metal–insulator phase boundary. As a result, the electrical properties of the Ca$_{1-x}$Sr$_x$IrO$_3$ films are found to be sensitive to $x$ and strain.

Numerous discoveries of fascinating physical phenomena relating to transition metal oxides (TMOs) have been reported, including a metal–insulator transition (MIT), high-temperature superconductivity, and colossal magnetoresistance.[1–5] Many of these phenomena can be understood in terms of Mott's physics, where the on-site Coulomb repulsion, $U$, and the bandwidth, $W$, are in competition with one another. Generally, when $U \leq W$, the system is metallic, but when $U \gg W$, it becomes insulating.[1,2]

In most 3$d$ or 4$d$ TMOs, $U \sim$ 3–5 eV, $W \sim$ 3–5 eV, and the crystal field splitting is typically 2–3 eV. The spin-orbit (SO) coupling is typically of the order of 0.1 eV for most 3$d$ or 4$d$ TMOs, and therefore is usually neglected when describing the physical properties.[2] On the other hand, the situation is quite different for 5$d$ TMOs. It is predicted that $W$ ($U$) of 5$d$ TMOs is larger (smaller) than that of 3$d$ and 4$d$ TMOs, as the wavefunctions are more spatially extended. Therefore, according to the Mott's physics, we can expect that most 5$d$ TMOs should be metallic.[6,7] However, some 5$d$ TMOs, such as $Sr_2IrO_4$, $Sr_3Ir_2O_7$, and $Ba_2NaOsO_6$, are known to have insulating ground states.[8–11] To solve such controversy, previous workers pointed out the importance of the role of the SO coupling.[12–15] As the strength of the SO coupling is proportional to $Z^4$ (where $Z$ is the atomic number), it can be as large as 0.3–0.5 eV in 5$d$ TMOs,[16] making the magnitude of the SO coupling comparable to $U$ or $W$.

Recently, we demonstrated that SO coupling should play a significant role in the physical properties of 5$d$ $Sr_2IrO_4$.[17] It was found that the insulating ground state could be described more accurately by considering an effective total angular momentum $J_{eff} = 1/2$ ($J_{eff,1/2}$) state in the strong SO coupling limit, rather than the well-known spin $S = 1/2$ state for conventional Mott insulators. The SO coupling induces the formation of $J_{eff,1/2}$ and $J_{eff,3/2}$ bands, which are occupied by five electrons. The $J_{eff,1/2}$ bands can be very narrow due to a reduced hopping integral caused by the isotropic orbital and mixed spin characteristics. Therefore, even when $U$ is small, the $J_{\text{eff},1/2}$ bands can be split into a lower Hubbard band (LHB) and an upper Hubbard band (UHB), opening a Mott gap. This study was extended to the Ruddlesden–Popper series of $Sr_{n+1}Ir_nO_{3n+1}$ ($n$ = 1, 2, and ∞),[18] and it was found that $W$ was expected to become larger when the number of neighboring Ir atoms, z, increases. Experimentally, it was found that the layered perovskite $Sr_2IrO_4$ (z = 4) and $Sr_3Ir_2O_7$ (z = 5) are in insulating states, but that the $SrIrO_3$ (z = 6) is in a metallic state. Moreover, electrodynamic studies of $SrIrO_3$ showed that it should be in a correlated metallic state, indicating that it is quite close to the metal–insulator (MI) phase boundary.[18]

Considering the dimensionality-controlled MIT in $Sr_{n+1}Ir_nO_{3n+1}$, perovskite $Ca_{1-x}Sr_xIrO_3$ should be a very intriguing material system. As the ionic size of the $A$-site ion becomes smaller than that of Sr ion, the distortion angle of the $IrO_6$ octahedra should increase. Such a structural change could result in a decrease of $W$, (Ref. 19) which would provide us with an opportunity to study the $W$-controlled MIT in 5$d$ TMOs. However, bulk $CaIrO_3$ is known to have a post-perovskite crystal structure, and $SrIrO_3$ to have a hexagonal crystal structure.[20,21] Therefore in order to investigate $W$-controlled MIT and the role of the SO coupling in the 5$d$ TMOs, it is highly desirable to fabricate perovskite $Ca_{1-x}Sr_xIrO_3$ phases.

In this paper, we report growth of perovskite $Ca_{1-x}Sr_xIrO_3$ ($x$ = 0, 0.5, and 1) thin films and their electronic and structural properties. We fabricated meta-stable perovskite $Ca_{1-x}Sr_xIrO_3$ thin films using epi-stabilization techniques[22] and investigated the electronic structure using transport measurements, optical spectroscopy, and first-principles calculations. We found that perovskite $Ca_{1-x}Sr_xIrO_3$ is located very close to an MI phase boundary, so the electronic properties are quite sensitive to external perturbations such as $x$ and strain. The first-principles calculations indeed showed that the SO coupling should play an important role in putting the iridates close to an MI phase boundary.

We used pulsed laser deposition to fabricate high-quality perovskite $CaIrO_3$, $Ca_{0.5}Sr_{0.5}IrO_3$, and $SrIrO_3$ thin films. We used an epi-stabilization technique[22] to fabricate the perovskite form by depositing them on to single crystalline $GdScO_3$(110) substrates. Note that the atomic arrangements of the substrate surfaces form a rectangular network, which is very close to the (100) surface of the perovskite structure. By maintaining the coherent film–substrate interface and minimizing the surface energy, meta-stable perovskite $CaIrO_3$, $Ca_{0.5}Sr_{0.5}IrO_3$, and $SrIrO_3$ phases were formed. Using high-resolution X-ray diffraction (HRXRD) measurements, we confirmed that all of the thin films were grown epitaxially. The thickness of the films was approximately 40 nm.

We measured the temperature-dependent resistivity, $\rho(T)$, using a four-point probe method. We also obtained near-normal-incidence reflectance and transmittance spectra of the thin films in the photon energy range 0.2–3.0 eV. We used a Fourier transform infrared spectrometer (Bruker IFS66v/S) for the range 0.2–1.2 eV and a grating-type spectrophotometer (CARY 5G) at 0.4–3.0 eV. We could determine the in-plane optical conductivity, $\sigma(\omega)$, of the perovskite $CaIrO_3$, $Ca_{0.5}Sr_{0.5}IrO_3$, and $SrIrO_3$ thin films from the transmittance and reflectance spectra by solving the Fresnel equations numerically.[23]

Figure 1(a) shows an XRD $\theta$–$2\theta$ pattern for the $CaIrO_3$ film on $GdScO_3$(110) substrate. The strongest sharp peaks are Bragg reflections from the $GdScO_3$(110) substrate. The pattern shows pure (00$l$)-oriented perovskite $CaIrO_3$ reflections, with no trace of impurities or additional phases. The calculated $c$-axis lattice constant is around 3.872Å. For comparison, the peak position of post-perovskite $CaIrO_3$ phase is also shown. Fig. 1(b) shows X-ray reciprocal space mapping (X-RSM) to a pseudo-cubic reciprocal lattice unit (r.l.u.) of GdScO$_3$ substrate. The X-RSM clearly shows that the pseudo-cubic (-103)$_C$ reflection of perovskite CaIrO$_3$ phase is on the same pseudo-cubic reciprocal plane of GdScO$_3$. As shown in the inset of Fig. 1(a), the $\varphi$-scans of the CaIrO$_3$(-103)$_C$ reveal a four-fold symmetry, having the same peak positions as GdScO$_3$(-103)$_C$. The X-RSM and $\varphi$-scans indicate that the perovskite CaIrO$_3$ film was deposited epitaxially. Moreover, the X-RSM data show that the perovskite CaIrO$_3$ film is almost fully strained to the substrate. We also investigated the structural properties of Ca$_{0.5}$Sr$_{0.5}$IrO$_3$ and SrIrO$_3$ films, and found that they also have similar high-quality perovskite phases (not shown).

Figure 2 shows $\rho(T)$ for the perovskite CaIrO$_3$, Ca$_{0.5}$Sr$_{0.5}$IrO$_3$, and SrIrO$_3$ thin films. Interestingly, the temperature-dependence of $\rho$ is quite weak for all Ca$_{1-x}$Sr$_x$IrO$_3$ compounds, and d$\rho$/d$T$ changes its sign from positive to negative as $x$ decreases. SrIrO$_3$ ($x = 1$) shows metallic behavior (d$\rho$/d$T > 0$) and has a resistivity of just less than 10$^{-3}$ $\Omega$cm, which corresponds to the Mott minimum metallic conductivity.[1] Note that a previous optical study showed that SrIrO$_3$ is a correlated metal near the MI phase boundary,[18] which is consistent with this transport measurement result. Ca$_{0.5}$Sr$_{0.5}$IrO$_3$ ($x = 0.5$) also shows a nearly metallic behavior (d$\rho$/d$T > 0$) and has a $\rho$ slightly larger than that of SrIrO$_3$, laying at the Mott boundary. The value of $\rho$ for CaIrO$_3$ ($x = 0$) is somewhat above the Mott boundary and exhibits insulator-like behavior (d$\rho$/d$T < 0$). However, $\rho(T)$ does not diverge at very low temperatures, which suggests that it might have semi-metallic behavior. Considering the highly metallic characters of 4$d$ perovskite Ca$_{1-x}$Sr$_x$RuO$_3$ and Ca$_{1-x}$Sr$_x$RhO$_3$ compounds, the 5$d$ Ca$_{1-x}$Sr$_x$IrO$_3$ films seem to constitute a unique system, which is located near the borderline of MIT.

Figures 3(a) and 3(b) show $\sigma(\omega)$ of the perovskite Ca$_{1-x}$Sr$_x$IrO$_3$ films at room temperature. Most $d$–$d$ transitions between the Ir 5$d$ orbital states were found to be located below 2 eV.[18]

Values of $\sigma(\omega)$ below 0.5 eV show clear changes as $x$ decreases, which is consistent with the transport data. For SrIrO$_3$, a distinct Drude-like response appears due to the free charge carriers. For Ca$_{0.5}$Sr$_{0.5}$IrO$_3$, the Drude-like feature is slightly suppressed; however, it still dominates the low energy optical response. On the other hand, for CaIrO$_3$, a peak structure develops as $\omega$ approaches ~0.2 eV from the high frequency side. In order to ensure that a sharp peak structure and not a Drude peak is present, we independently measured the reflectance spectra of far-infrared energy region (6-80 meV)[24] and obtained $\sigma(\omega)$ below 80 meV.[25] We found that CaIrO$_3$ has a strong suppression of $\sigma(\omega)$ in the far infrared region. These spectral changes suggest that an MIT–like transition occurs in Ca$_{1-x}$Sr$_x$IrO$_3$ when $x$ increases.

The width of the sharp CaIrO$_3$ peak at ~0.2 eV is much narrower than those of the correlation-induced peaks in other 3$d$ and 4$d$ TMOs. We denote this sharp peak with $\alpha$ and a broad peak around 0.75 eV with $\beta$. The peculiar double-peak structure, which is quite scarce for typical Mott insulators, has been also observed in Sr$_2$IrO$_4$ and Sr$_3$Ir$_2$O$_7$.[18] These peaks originated from the characteristic feature of the SO-coupling-triggered $J_{\text{eff},1/2}$ Hubbard bands and $J_{\text{eff},3/2}$ bands in the perovskite structure. For comparison, we plotted $\sigma(\omega)$ of Sr$_2$IrO$_4$ and Sr$_3$Ir$_2$O$_7$, shown in Fig 3(b). According to Fermi's golden rule, the width of an absorption peak should reflect $W$ of the initial and final bands. The width of peak $\alpha$ for CaIrO$_3$ from the Lorentz oscillator model fitting is about 0.42 eV, which is close to that of Sr$_3$Ir$_2$O$_7$ (~0.45 eV) and larger than that of Sr$_2$IrO$_4$ (~0.27 eV).[18] The results indicate that $W$ of $J_{\text{eff},1/2}$ bands in CaIrO$_3$ is similar to that of $J_{\text{eff},1/2}$ bands in Sr$_3$Ir$_2$O$_7$, which was smaller than that of $J_{\text{eff},1/2}$ bands in SrIrO$_3$ due to the reduced dimensionality.

To gain more insight into the electronic structure of perovskite CaIrO$_3$, we performed local density approximation (LDA) + $U$ calculations with SO coupling included. For the first-principle analysis, we used the linear combination of pseudo-atomic orbital (LCPAO)-based code OpenMX, in which a fully relativistic *j*-dependent pseudo-potential and LDA + *U* scheme is implemented. We used double *s* and *p*, and single *d* pseudo-atomic-orbital calculations for all of the atoms, with cut-off radii of 7.0 a.u. for Ca and Ir and 5.0 a.u. for oxygen. For the *k*-grid integration, we used (10 × 10 × 7) *k*-space points over the first Brillouin zone and a 400-Ry energy cut-off for the real-space numerical integration and solution of Poisson's equation. For the lattice structure, we used the lattice constants given by the XRD analysis as a starting point and carried out full lattice relaxation up to $5.0 \times 10^{-4}$ Hartree/Å of force criterion. For the LDA + SO + *U* calculations, *U* = 2.0 eV was used.

Figures 3(c) and 3(d) show the calculated band structures of $CaIrO_3$ with the LDA and LDA + SO + *U* calculations, respectively. In the energy region between -2.5 and 0.5 eV, the Ir 5*d* $t_{2g}$ states were the main contributors. The LDA result in Fig. 3(c) yields a metallic ground state with complex $t_{2g}$ bands crossing the Fermi energy ($E_F$). When the SO coupling is included, the band structure changes remarkably: the bands crossing $E_F$ are split off due to formation of the $J_{\text{eff},1/2}$ and $J_{\text{eff},3/2}$ bands. As *U* becomes involved in the system, the upper part of the $J_{\text{eff},1/2}$ band slightly shifts to higher energies, and the lower part of the $J_{\text{eff},1/2}$ and $J_{\text{eff},3/2}$ bands slightly shift to lower energies. The light and dark lines in Fig. 3(d) represent the $J_{\text{eff},1/2}$ and $J_{\text{eff},3/2}$ bands, respectively.

We assigned the optical transition peaks shown in Fig. 3(b) according to calculated data. The transitions from the lower $J_{\text{eff},1/2}$ to upper $J_{\text{eff},1/2}$ bands and from the $J_{\text{eff},3/2}$ bands to the upper $J_{\text{eff},1/2}$ bands result in the peaks *α* and *β*, respectively. Because the narrow $J_{\text{eff},1/2}$ bands are located near $E_F$, the increase in *W* could easily induce the density of states at the $E_F$. This effect should be reflected as the broadening of peak $\alpha$ and finally results in the coherent Drude response in $\sigma(\omega)$.

It should be noted that, although a gap is opened, the conduction and valence bands still touch $E_F$, resulting in a small hole and electron pocket, as shown in Fig. 3(d). This phenomenon is similar to semi-metallic behavior. In the transport data, the overall behavior of $\rho(T)$ is insulator-like, but does not diverge at low temperatures, which is also consistent with the semi-metallic character. On the other hand, the optical transition is quite sensitive to the direct transition, so semi-metallic character would be difficult to observe. This semi-metallic behavior indicates that $CaIrO_3$ is positioned between metallic and insulating phases during the process of the band dispersion lifting by $IrO_6$ distortion.

From the experimental and theoretical results, we can conclude that the strong SO coupling pushes the $Ca_{1-x}Sr_xIrO_3$ system into the vicinity of an MIT by inducing the formation of $J_{eff}$ bands. Due to this Mott instability, it is expected that the electrical ground state could be easily tuned by subtle external perturbations, such as changes in the lattice parameters. To demonstrate such a high sensitivity, we controlled the lattice parameters of $Ca_{0.5}Sr_{0.5}IrO_3$, which can affect $W$, by epitaxially growing the films on different substrates. That is, we deposited $Ca_{0.5}Sr_{0.5}IrO_3$ on a $SrTiO_3$ substrate, which has a smaller lattice constant than $GdScO_3$, resulting in compressive strain. Figure 4(a) shows XRD $\theta$–$2\theta$ patterns for the $Ca_{0.5}Sr_{0.5}IrO_3$ film on $GdScO_3(110)$ and $SrTiO_3(001)$ substrates. Because the lattice parameter of $Ca_{0.5}Sr_{0.5}IrO_3$ is between $GdScO_3$ and $SrTiO_3$, the $Ca_{0.5}Sr_{0.5}IrO_3$ film is under a tensile strain when it is grown on $GdScO_3$ and a compressive strain when it is grown on $SrTiO_3$. Figure 4(b) shows $\rho(T)$ for the $Ca_{0.5}Sr_{0.5}IrO_3$ thin films on $GdScO_3$ and $SrTiO_3$ substrates. The film that was grown on $SrTiO_3$ shows insulator-like behavior, whereas the film on $GdScO_3$ shows nearly metallic behavior, which suggests that the compressive strain of SrTiO$_3$ substrate enhances the in-plane distortion that makes $W$ of the $J_{\text{eff},1/2}$ bands decrease. This result demonstrates that the Ca$_{1-x}$Sr$_x$IrO$_3$ system is indeed very close to the Mott instability due to the strong SO coupling of the 5$d$ Ir ion.

In summary, we successfully fabricated meta-stable perovskite Ca$_{1-x}$Sr$_x$IrO$_3$ ($x$ = 0, 0.5, and 1) thin films and observed that the electronic properties are sensitive to changes in $x$ and strain. Using optical spectroscopy and first-principles calculations, we demonstrated that strong spin-orbit coupling results in the perovskite Ca$_{1-x}$Sr$_x$IrO$_3$ being located very close to a metal–insulator phase boundary due to the formation of effective total angular momentum $J_{\text{eff}}$ = 1/2 and 3/2 states. This work provides the further advancement in understanding the underlying physics of $J_{\text{eff}}$ state in 5$d$ transition metal oxides and manipulating it by changing chemical pressure and strain state.

This research was supported by the Basic Science Research Program through the National Research Foundation of Korea (NRF) funded by the Ministry of Education, Science and Technology (No. 2009-0080567).

**Figure captions**

Figure 1. (color online) (a) X-ray $\theta$–$2\theta$ scan of the CaIrO$_3$ film on GdScO$_3$(110) substrate. The triangles indicate the peak positions of the post-perovskite CaIrO$_3$ phase. (b) X-ray reciprocal space mapping around the (-103)$_C$ Bragg reflection from the GdScO$_3$ substrate and CaIrO$_3$ film peaks. The inset of (a) shows the $\varphi$-scans of the (-103)$_C$ peaks for CaIrO$_3$ film and GdScO$_3$ substrate, which demonstrates that we have epitaxial growth of the perovskite CaIrO$_3$ phase.

Figure 2. (color online) Temperature-dependent resistivity, $\rho(T)$, of CaIrO$_3$, Ca$_{0.5}$Sr$_{0.5}$IrO$_3$, and SrIrO$_3$ thin films.

Figure 3. (color online) Optical conductivity spectra, $\sigma(\omega)$, of (a) SrIrO$_3$, Ca$_{0.5}$Sr$_{0.5}$IrO$_3$, and (b) CaIrO$_3$ thin films. In (b), $\sigma(\omega)$ of Sr$_2$IrO$_4$ and Sr$_3$Ir$_2$O$_7$ (from Ref. 18) are also shown for comparison. Theoretical dispersion relations of CaIrO$_3$ from (c) LDA and (d) LDA + SO + $U$ calculations.

Figure 4. (color online) (a) X-ray $\theta$–$2\theta$ scans and (b) $\rho(T)$ of the Ca$_{0.5}$Sr$_{0.5}$IrO$_3$ films on GdScO$_3$(110) and SrTiO$_3$(001) substrates.

[25] In the far-infrared region, the strong phonon signal of $GdScO_3$ substrate barely allows the transmission of light. Hence, we tried to extract the $\sigma(\omega)$ from the Lorentz oscillator model fit of reflectance spectrum. However, due to its low symmetry, the phonon modes of $GdScO_3$ are too complicated to extract pure optical response of the film. As $SrTiO_3$ has rather simple phonon mode, we measured the reflectance of $CaIrO_3$ film on $SrTiO_3$ substrate. For the $CaIrO_3$ film on $SrTiO_3$, the perovskite $CaIrO_3$ phase was also confirmed by HRXRD measurements and an insulator-like temperature-dependent resistivity curve, which has slightly larger value than that of $CaIrO_3$/$GdScO_3$, was obtained from transport measurement.

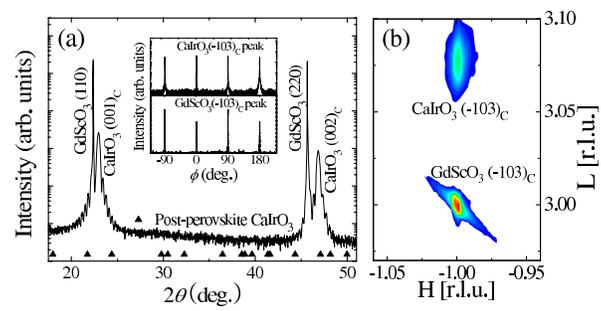

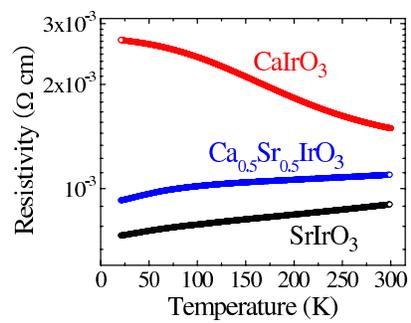

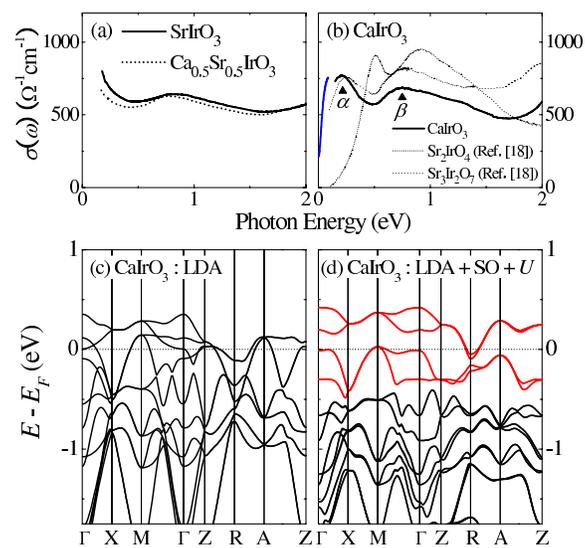

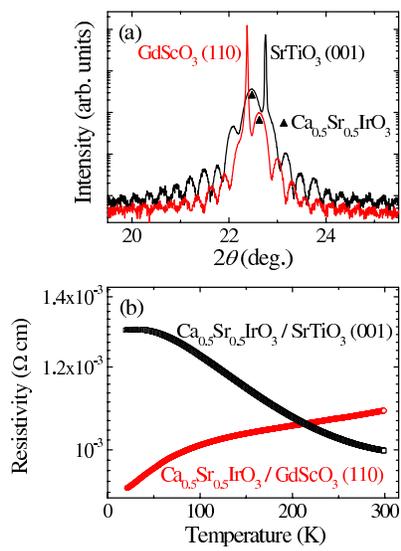

Figure 4   BLJ1104   17NOV2009